\documentclass[prb,twocolumn,showpacs,preprintnumbers]{revtex4-1}
\usepackage{amsmath,bm,graphicx,hyperref}

\renewcommand\d{\partial}
\newcommand\grad{\bm{\nabla}}
\newcommand\+{\dagger}
\newcommand\<{\langle}
\renewcommand\>{\rangle}
\newcommand\0{{\bm{0}}}
\newcommand\e{{\bm{e}}}
\renewcommand\r{{\bm{r}}}
\newcommand\x{{\bm{x}}}
\newcommand\y{{\bm{y}}}
\newcommand\z{{\bm{z}}}
\renewcommand\k{{\bm{k}}}
\newcommand\p{{\bm{p}}}
\newcommand\q{{\bm{q}}}
\newcommand\ep{\varepsilon_\p}
\newcommand\eq{\varepsilon_\q}
\newcommand\kB{k_\mathrm{B}}
\newcommand\tr{\mathrm{tr}}
\newcommand\Tr{\mathrm{Tr}}
\renewcommand\Re{\mathrm{Re}}
\renewcommand\Im{\mathrm{Im}}

\begin{document}
\preprint{LA-UR-13-20791}

\title{Electron spin resonance in a dilute magnon gas\\
as a probe of magnon scattering resonances}

\author{Yusuke Nishida}
\affiliation{Theoretical Division, Los Alamos National Laboratory,
Los Alamos, New Mexico 87545, USA}
\affiliation{Department of Physics, Tokyo Institute of Technology,
Ookayama, Meguro, Tokyo 152-8551, Japan}

\date{February 2013}

\begin{abstract}
 We study the electron spin resonance in a dilute magnon gas that is
 realized in a ferromagnetic spin system at low temperature.  A quantum
 cluster expansion is developed to show that the frequency shift of the
 single-magnon peak changes its sign and the linewidth reaches its
 maximum across a scattering resonance between magnons.  Such
 characteristic behaviors are universal and can be used to
 experimentally locate the two-magnon resonance when an external
 parameter such as pressure is varied.  Future achievement of the
 two-magnon resonance may have an impact comparable to the Feshbach
 resonance in ultracold atoms and will open up a rich variety of
 strongly correlated physics such as the recently proposed Efimov effect
 in quantum magnets.  We also suggest how the emergence of an Efimov
 state of three magnons and its binding energy may be observed with the
 electron spin resonance.
\end{abstract}


\maketitle

\section{Introduction}
Scattering resonances play an important role in physics.  In the field
of ultracold atoms, Feshbach resonances induced with magnetic field are
used to control the interaction between atoms.\cite{Chin:2010}  This
unparalleled tunability of the interaction has been essential to the
recent remarkable advance of the field and led to experimental
realization of a rich variety of physics, such as the BCS-BEC crossover
in Fermi gases\cite{Regal:2004,Zwierlein:2004} and the Efimov effect in
Bose gases.\cite{Kraemer:2006}  These strongly correlated phenomena in
the vicinity of the scattering resonance are universal, i.e.,
independent of microscopic details.  Also, an atom loss peak caused by
the three-atom or atom-dimer resonance has been used as a signature of
the emergence of an Efimov trimer.\cite{Kraemer:2006,Knoop:2009}

Turning to the field of condensed matter, the scattering resonance can
be induced between collective excitations in ferromagnetic spin systems
(magnons), for example, by tuning the exchange
coupling.\cite{Nishida:2013}  Because the ferromagnetic coupling is
sensitive to the structure of ions,\cite{Goodenough:1963} it is
experimentally possible to vary the exchange coupling significantly with
pressure\cite{Kawamoto:2001} and bring the system close to the
two-magnon resonance.  Future achievement of the two-magnon resonance
may have an impact comparable to the Feshbach resonance in ultracold
atoms and will open up a rich variety of strongly correlated physics
such as the recently proposed Efimov effect in quantum
magnets.\cite{Nishida:2013}  But, what is an experimental signature of
the two-magnon resonance?

In this paper, we study the electron spin resonance (ESR) in a dilute
magnon gas as a probe of magnon scattering resonances.  ESR is a
powerful experimental technique to investigate the dynamics of
interacting spin systems,\cite{Katsumata:2000,Ajiro:2003} while
theoretical calculations of its spectrum are in general challenging and
thus often suffer from some limitations.  For example, the previous
approaches developed by Kubo-Tomita\cite{Kubo:1954} and
Mori-Kawasaki\cite{Mori:1962} are based on perturbations in terms of
small spin anisotropies,\cite{Oshikawa:2002} which are actually of no
use for our purpose because magnon scattering resonances take place at
large spin anisotropies.\cite{Nishida:2013}  Therefore, a new approach
must be developed where spin anisotropies can be treated
nonperturbatively.  One such approach is a quantum cluster (or virial)
expansion which uses a fugacity $z\equiv e^{-\beta\Delta}$ as a small
parameter to perform a systematic expansion.\cite{Landau-Lifshitz:1980}
Here $\Delta$ is a single-particle excitation energy,
$\beta\equiv1/(\kB T)$ the inverse temperature, and the system is
assumed to be so dilute that $z\ll1$, but importantly, no assumptions
are needed on the form of the interaction.

We first demonstrate this quantum cluster expansion in
Sec.~\ref{sec:boson} by computing a single-particle spectral function in
a dilute Bose gas, which serves as the universal formula for ESR at low
temperature and can be used to extract the scattering length between
magnons.  Then in Sec.~\ref{sec:magnon}, we apply the same method to ESR
in a dilute magnon gas and show that the frequency shift of the
single-magnon peak changes its sign and the linewidth reaches its
maximum across a scattering resonance between magnons.  Such
characteristic behaviors are universal and can be used to experimentally
locate the two-magnon resonance when an external parameter such as
pressure is varied.  Finally, Sec.~\ref{sec:summary} is devoted to the
conclusion of this paper and discussion where we also suggest how the
emergence of an Efimov state of three magnons and its binding energy may
be observed with ESR.  Below we set $\hbar=\kB=1$ and the unspecified
range of integration is assumed to be from $-\infty$ to $\infty$.

\section{Single-particle spectral function in a dilute Bose gas \label{sec:boson}}
Before we work on ESR in a dilute magnon gas, it is instructive to
demonstrate the quantum cluster expansion in a dilute Bose gas, which is
described by a Hamiltonian
\begin{align}
 H = \int\!d\r\,\psi_\r^\+(\varepsilon_{-i\grad}+\Delta)\psi_\r
 - \frac{g}2\int\!d\r\,\psi_\r^\+\psi_\r^\+\psi_\r\psi_\r.
\end{align}
Here $\varepsilon_\p=\p^2/(2m)$ is the single-particle dispersion
relation, $\mu=-\Delta<0$ is the chemical potential, and the bare
coupling constant $g$ is related to the $s$-wave scattering length $a_s$
by $1/g=m\Lambda/(2\pi^2)-m/(4\pi a_s)$ with $\Lambda$ being a momentum
cutoff.  Our purpose in this section is to compute the single-particle
spectral function
\begin{align}\label{eq:boson_spectrum}
 A(\omega,\p) = -2\,\Im\,G(\omega+i0^+,\p),
\end{align}
where $G(i\omega_n,\p)$ with $\omega_n\equiv2\pi n/\beta$ being the
Matsubara frequency is the Fourier transform of the imaginary-time
propagator
\begin{align}\label{eq:propagator-freq}
 G(i\omega_n,\p) = -\int_0^\beta\!d\tau\!\int\!d\r\,
 e^{i\omega_n\tau-i\p\cdot\r}\,\<\psi_\r(\tau)\psi_\0^\+(0)\>.
\end{align}
As we will show in the next section, $A(\omega,\0)$ serves as the
universal formula for ESR at low temperature and can be used to extract
the scattering length between magnons.

The imaginary-time propagator is defined by
\begin{align}\label{eq:propagator-time}
 -\<\psi_\r(\tau)\psi_\0^\+(0)\>
 = -\frac1Z\,\Tr[e^{-\beta H}\psi_\r(\tau)\psi_\0^\+(0)],
\end{align}
where $Z$ is the grand canonical partition function:
$Z=\Tr[e^{-\beta H}]$.  The systematic expansion over the fugacity
$z=e^{-\beta\Delta}\ll1$ can be developed by writing the grand canonical
trace as a sum over canonical traces with fixed particle number $N$:
$\Tr[\,\cdot\,]=\sum_{N=0}^\infty\tr_N[\,\cdot\,]$.  Because
$\tr_N[e^{-\beta H}]\propto z^N$, we obtain
\begin{align}\label{eq:boson_partition}
 Z = 1 + V\frac{z}{\lambda^3} + O(z^2),
\end{align}
where $V$ is the system volume and $\lambda\equiv\sqrt{2\pi\beta/m}$ is
the thermal de Broglie wavelength.  Accordingly, the particle number
density is found to be
\begin{align}
 n = \frac1{V\beta}\frac{\d\ln Z}{\d\mu} = \frac{z}{\lambda^3} + O(z^2).
\end{align}
Therefore, the quantum cluster expansion is valid when the system is so
dilute that the mean interparticle distance is much larger than the
thermal de Broglie wavelength: $n^{-1/3}\gg\lambda$.

The numerator in Eq.~(\ref{eq:propagator-time}) can be expanded over $z$
in the same way.  By denoting the Fourier transform of each term as
\begin{align}
 G_N(i\omega_n,\p) &\equiv -\int_0^\beta\!d\tau\!\int\!d\r\,
 e^{i\omega_n\tau-i\p\cdot\r} \notag\\
 &\times \tr_N[e^{-\beta H}\psi_\r(\tau)\psi_\0^\+(0)] \sim O(z^N),
\end{align}
the leading term is easily evaluated as
\begin{align}\label{eq:propagator_0}
 G_0(i\omega_n,\p) = \frac{1-e^{-\beta\ep}z}{i\omega_n-\ep-\Delta}.
\end{align}
On the other hand, after a straightforward calculation, the
next-to-leading term is evaluated as
\begin{align}\label{eq:propagator_1}
 & G_1(i\omega_n,\p)
 = \frac{V/\lambda^3+e^{-\beta\ep}}{i\omega_n-\ep-\Delta}\,z \notag\\
 &+ \int\!\frac{d\q}{(2\pi)^3}
 \frac{F(i\omega_n,\p;\q)}{(i\omega_n-\ep-\Delta)^2}e^{-\beta\eq}z + O(z^2),
\end{align}
where
\begin{align}\label{eq:boson_amplitude}
 F(i\omega_n,\p;\q) \equiv \frac{8\pi/m}
 {1/a_s-\sqrt{-m(i\omega_n-\Delta+\eq-\varepsilon_{\p+\q}/2)}}
\end{align}
is the forward scattering amplitude between a particle with
energy-momentum $(i\omega_n-\Delta,\p)$ and an on-shell particle with
momentum $\q$.  Then, by writing the Fourier transform of the
imaginary-time propagator (\ref{eq:propagator-freq}) in the standard
form
\begin{align}
 G(i\omega_n,\p) = \frac1{i\omega_n-\ep-\Delta-\Sigma(i\omega_n,\p)}
\end{align}
and comparing it with its systematic expansion obtained in
Eqs.~(\ref{eq:boson_partition}), (\ref{eq:propagator_0}), and
(\ref{eq:propagator_1}), we find that the self-energy
$\Sigma(i\omega_n,\p)$ must have the following quantum cluster
expansion:
\begin{align}
 \Sigma(i\omega_n,\p) = z\int\!\frac{d\q}{(2\pi)^3}
 F(i\omega_n,\p;\q)\,e^{-\beta\eq} + O(z^2).
\end{align}
Therefore, the self-energy at $O(z)$ is determined only by the
two-particle physics, i.e., binary collisions with thermally excited
particles.

\begin{figure}[t]
 \includegraphics[width=0.9\columnwidth,clip]{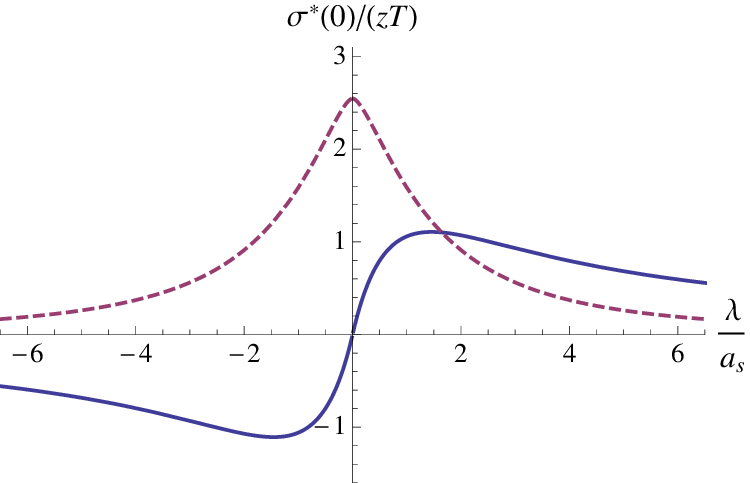}
 \caption{Real and imaginary parts of $\sigma^*(\0)/(zT)$ in
 Eq.~(\ref{eq:boson_on-shell}) as functions of $\lambda/a_s$,
 represented by solid and dashed curves, respectively.
 \label{fig:boson}}
\end{figure}

The resulting single-particle spectral function
(\ref{eq:boson_spectrum}) at $O(z^0)$ is simply a delta function located
at $\omega=\ep+\Delta$ corresponding to the single-particle energy.  By
including the self-energy correction
$\Sigma(i\omega_n,\p)\sim O(z)\ll1$,  it becomes a sharp peak whose line
shape within the accuracy up to $O(z)$ is described by the Lorentzian
\begin{align}\label{eq:boson_result}
 A_\mathrm{peak}(\omega,\p) \approx \frac{-2\,\Im\,\sigma(\p)}
 {[\omega-\ep-\Delta-\Re\,\sigma(\p)]^2+[\Im\,\sigma(\p)]^2},
\end{align}
where we introduced the on-shell self-energy
\begin{align}\label{eq:boson_on-shell}
 \sigma(\p) &\equiv \Sigma(\ep+\Delta+i0^+,\p) \notag\\
 &= z\int\!\frac{d\q}{(2\pi)^3}\frac{8\pi/m}{1/a_s+i|\p-\q|/2}\,e^{-\beta\eq}.
\end{align}
Therefore, the energy shift and the decay width of a particle with
momentum $\p$ in a dilute Bose gas are given by the real and imaginary
parts of $\sigma^*(\p)$, respectively.  The corresponding normalized
quantity $\sigma^*(\p)/(zT)\sim O(1)$ at $\p=\0$ is plotted in
Fig.~\ref{fig:boson} as a function of $\lambda/a_s$.  We find that the
energy shift changes its sign and the decay width reaches its maximum
across the scattering resonance at $a_s\to\infty$.  Such characteristic
behaviors are universal and shared by ESR at low temperature.

\begin{figure}[t]
 \includegraphics[width=0.9\columnwidth,clip]{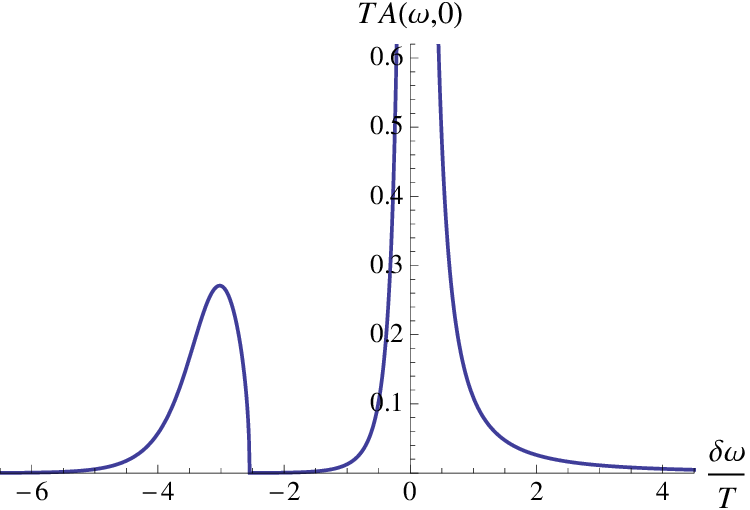}
 \caption{Single-particle spectral function $A(\omega,\0)$ for $z=0.1$
 and $\lambda/a_s=4$ as a function of $\delta\omega\equiv\omega-\Delta$.
 The single-particle peak goes up to $TA(\omega,\0)\simeq51$ and there
 is a threshold singularity at the two-particle binding energy
 $E_2/T=-(\lambda/a_s)^2/(2\pi)$.  \label{fig:bound}}
\end{figure}

Besides the single-particle peak at $\omega=\ep+\Delta+\Re\,\sigma(\p)$,
there exists an additional structure in $A(\omega,\p)$ at
$\omega<\ep+\Delta+E_2$ when two particles form a bound state with
binding energy $E_2=-1/(ma_s^2)<0$.  This structure is due to the pole
of the two-particle scattering amplitude (\ref{eq:boson_amplitude}) and
shows a square-root threshold singularity at $\omega=\ep+\Delta+E_2$ in
the single-particle spectral function as
$A(\omega,\p)\sim z\sqrt{\ep+\Delta+E_2-\omega}$.  This feature at
$\p=\0$ is demonstrated in Fig.~\ref{fig:bound} by choosing $z=0.1$ and
$\lambda/a_s=4$ as an example.  We also note that the single-particle
spectrum function of bosons is universal up to $O(z^2)$, i.e.,
independent of an ultraviolet cutoff.  The $O(z^2)$ term involves the
three-particle physics and thus depends on an additional parameter,
so-called Efimov parameter $\kappa_*$.\cite{Braaten:2006}  See the
discussion in Sec.~\ref{sec:summary} on how the emergence of an Efimov
state and its binding energy may appear in the expected behavior of
$A(\omega,\0)$ at $O(z^2)$.

\section{Electron spin resonance in a dilute magnon gas \label{sec:magnon}}

\subsection{Quantum cluster expansion}
We now apply the same method to ESR in a dilute magnon gas.  For
definiteness, we consider a spin-$S$ Heisenberg model with exchange
($J_z\neq J>0$) and single-ion ($D\neq0$) anisotropies, which is
described by
\begin{align}
 H &= -\frac12\sum_\r\sum_{\hat\e}\,
 (J\,S_\r^+S_{\r+\hat\e}^- + J_z\,S_\r^zS_{\r+\hat\e}^z) \notag\\
 &\quad - D\sum_\r(S_\r^z)^2 - B\sum_\r S_\r^z.
\end{align}
Here $\sum_{\hat\e}=\sum_{\hat\e=\pm\hat\x,\pm\hat\y,\pm\hat\z}$ is a
sum over six unit vectors of a simple cubic lattice and spin operators
$S_\r^{\pm}\equiv S_\r^x\pm iS_\r^y$ and $S_\r^z$ obey the usual
commutation relations: $[S_\r^+,S_{\r'}^-]=2S_\r^z\delta_{\r,\r'}$ and
$[S_\r^z,S_{\r'}^{\pm}]=\pm S_\r^\pm\delta_{\r,\r'}$.  ESR experiments
measure an absorption intensity of electromagnetic radiation polarized
perpendicular to the magnetic field axis.  Within the linear response
theory for a circular polarization, the absorption intensity normalized
by the system volume and the intensity of the incident radiation is
given by\cite{Oshikawa:2002,Brockmann:2012}
\begin{align}\label{eq:magnon_spectrum}
 I(\omega) = \frac\omega2\,\Im\,\chi(\omega+i0^+,\0),
\end{align}
where $\chi(i\omega_n,\p)$ is the Fourier transform of the
imaginary-time susceptibility
\begin{align}\label{eq:susceptibility-freq}
 \chi(i\omega_n,\p) = \int_0^\beta\!d\tau\sum_\r\,
 e^{i\omega_n\tau-i\p\cdot\r}\,\<S_\r^-(\tau)S_\0^+(0)\>.
\end{align}
While the ESR experiments can measure the spectrum only at zero
momentum, we develop the formulation for general $\p$ and set $\p=\0$ at
the end.

The ground state for a sufficiently large magnetic field $B<0$ is a
fully polarized state with all spins pointing downwards:
$S_\r^z|0\>=-S|0\>$ and $S_\r^-|0\>=0$.  Accordingly, we redefine the
Hamiltonian to absorb the ground state energy
$E_0=-\sum_\r(3J_zS^2+DS^2-BS)$ so that $H|0\>=0$.  Because of the U(1)
symmetry under rotation $S_\r^\pm\to e^{\pm i\theta}S_\r^\pm$, the
magnetization relative to the ground state $\delta M\equiv\<S_\r^z\>+S$
is a conserved quantity which corresponds to a particle number density
of magnons.  Then at low temperature, magnons are thermally excited and
thus the system becomes a dilute magnon gas.  A single magnon has the
dispersion relation $\ep=SJ\sum_{\hat\e}[1-\cos(\p\cdot\hat\e)]$ with
the excitation energy $\Delta=-6SJ+6SJ_z+2SD-D-B$.  As long as the
fugacity is small, $z=e^{-\beta\Delta}\ll1$, the quantum cluster
expansion can be developed for the dilute magnon gas similarly to the
previous dilute Bose gas. The imaginary-time susceptibility is defined
by
\begin{align}\label{eq:susceptibility-time}
 \<S_\r^-(\tau)S_\0^+(0)\> = \frac1Z\,\Tr[e^{-\beta H}S_\r^-(\tau)S_\0^+(0)],
\end{align}
where $Z$ is the grand canonical partition function:
$Z=\Tr[e^{-\beta H}]$.  By writing the grand canonical trace as a sum
over canonical traces with fixed magnon number $N$, we obtain
\begin{align}\label{eq:magnon_partition}
 Z = \sum_{N=0}^\infty\tr_N[e^{-\beta H}] = 1 + V\frac{z}{\rho^3} + O(z^2),
\end{align}
where we used $\tr_N[e^{-\beta H}]\propto z^N$ and introduced the
analog of the thermal de Broglie wavelength by
$\rho\equiv a\,e^{2\beta SJ}/I_0(2\beta SJ)$ with $a$ being the lattice
constant.  Accordingly, the relative magnetization is found to be
\begin{align}
 \delta M = \frac1{V\beta}\frac{\d\ln Z}{\d B} = \frac{z}{\rho^3} + O(z^2).
\end{align}

The numerator in Eq.~(\ref{eq:susceptibility-time}) can be expanded over
$z$ in the same way.  By denoting the Fourier transform of each term as
\begin{align}
 \chi_N(i\omega_n,\p) &\equiv \int_0^\beta\!d\tau\sum_\r\,
 e^{i\omega_n\tau-i\p\cdot\r} \notag\\
 &\times \tr_N[e^{-\beta H}S_\r^-(\tau)S_\0^+(0)] \sim O(z^N),
\end{align}
the leading term is easily evaluated as
\begin{align}\label{eq:susceptibility_0}
 \chi_0(i\omega_n,\p) = -2S\frac{1-e^{-\beta\ep}z}{i\omega_n-\ep-\Delta}.
\end{align}
On the other hand, after a straightforward calculation, the
next-to-leading term is evaluated as
\begin{align}\label{eq:susceptibility_1}
 & \chi_1(i\omega_n,\p)
 = -2S\,\frac{V/\rho^3+e^{-\beta\ep}}{i\omega_n-\ep-\Delta}\,z \notag\\
 &- 2S\int_{-\pi/a}^{\pi/a}\!\frac{d\q}{(2\pi/a)^3}
 \frac{\Gamma(i\omega_n,\p;\q)}{(i\omega_n-\ep-\Delta)^2}e^{-\beta\eq}z + O(z^2),
\end{align}
where
\begin{align}\label{eq:magnon_amplitude}
 \Gamma(i\omega_n,\p;\q) &\equiv \sum_{\hat\e}\,[J\cos(\tfrac{\p+\q}2\cdot\hat\e)
 - J_z\cos(\tfrac{\p-\q}2\cdot\hat\e)]\,\gamma(\hat\e) \notag\\
 &\quad - 2D\gamma(\0)
\end{align}
is the forward scattering amplitude between a magnon with
energy-momentum $(i\omega_n-\Delta,\p)$ and an on-shell magnon with
momentum $\q$.  Here the unknown function $\gamma(\r)=\gamma(-\r)$
implicitly depends on $(i\omega_n,\p;\q)$ and satisfies the
Lippmann-Schwinger equation
\begin{align}\label{eq:lippmann-schwinger}
 & \gamma(\r) = 2\cos(\tfrac{\p-\q}2\cdot\r)
 + \int_{-\pi/a}^{\pi/a}\!\frac{d\k}{(2\pi/a)^3}\cos(\k\cdot\r) \notag\\
 &\times \frac{\sum_{\hat\e}\,[J\cos(\tfrac{\p+\q}2\cdot\hat\e)
 - J_z\cos(\k\cdot\hat\e)]\,\gamma(\hat\e) - 2D\gamma(\0)}
 {i\omega_n-\Delta+\eq-\varepsilon_{(\p+\q)/2+\k}-\varepsilon_{(\p+\q)/2-\k}}.
\end{align}
By setting $\r=\hat\x,\,\hat\y,\,\hat\z$ and $\r=\0$, we obtain four
coupled equations to determine $\gamma(\hat\e)$ and $\gamma(\0)$
appearing in Eq.~(\ref{eq:magnon_amplitude}).

Then, by writing the Fourier transform of the imaginary-time
susceptibility (\ref{eq:susceptibility-freq}) in the standard form
\begin{align}
 \chi(i\omega_n,\p) = \frac{-2S}{i\omega_n-\ep-\Delta-\Xi(i\omega_n,\p)}
\end{align}
and comparing it with its systematic expansion obtained in
Eqs.~(\ref{eq:magnon_partition}), (\ref{eq:susceptibility_0}), and
(\ref{eq:susceptibility_1}), we find that the self-energy
$\Xi(i\omega_n,\p)$ must have the following quantum cluster expansion:
\begin{align}\label{eq:magnon_self-energy}
 \Xi(i\omega_n,\p) = z\int_{-\pi/a}^{\pi/a}\!\frac{d\q}{(2\pi/a)^3}
 \Gamma(i\omega_n,\p;\q)\,e^{-\beta\eq} + O(z^2).
\end{align}
Therefore, the self-energy at $O(z)$ is determined only by the
two-magnon physics, i.e., binary collisions with thermally excited
magnons.  The resulting ESR spectrum (\ref{eq:magnon_spectrum}) at
$O(z^0)$ is simply a delta function located at $\omega=\Delta$
corresponding to the single-magnon energy at $\p=\0$.  By including the
self-energy correction $\Xi(i\omega_n,\0)\sim O(z)\ll1$, it becomes a
sharp peak whose line shape within the accuracy up to $O(z)$ is
described by the Lorentzian
\begin{align}\label{eq:magnon_result}
 I_\mathrm{peak}(\omega) \approx \frac\omega2\frac{-2S\,\Im\,\xi(\0)}
 {[\omega-\Delta-\Re\,\xi(\0)]^2+[\Im\,\xi(\0)]^2},
\end{align}
where we introduced the on-shell self-energy:
$\xi(\p)\equiv\Xi(\ep+\Delta+i0^+,\p)$.  Therefore, the frequency shift
and the linewidth of the single-magnon peak are given by the real and
imaginary parts of $\xi^*(\0)$, respectively.\cite{footnote}  Also, when
two magnons form a bound state with binding energy $E_2<0$, the ESR
spectrum shows an additional structure at $\omega<\Delta+E_2$ similarly
to Fig.~\ref{fig:bound}, while we will not investigate it further.

\begin{figure*}[t]
 \hspace{11mm}\fbox{\scriptsize$\,S=1/2\,$}\hspace{41mm}
 \fbox{\scriptsize$\,S=1~\,(D=0)\,$}\hspace{37mm}\fbox{\scriptsize$\,S=1~\,(J_z=J)\,$}\smallskip\medskip\\
 \rotatebox{90}{\small\hspace{13.5mm}$a_s${\scriptsize/}$a$}\hfill\hspace{0mm}
 \includegraphics[width=0.318\textwidth,clip]{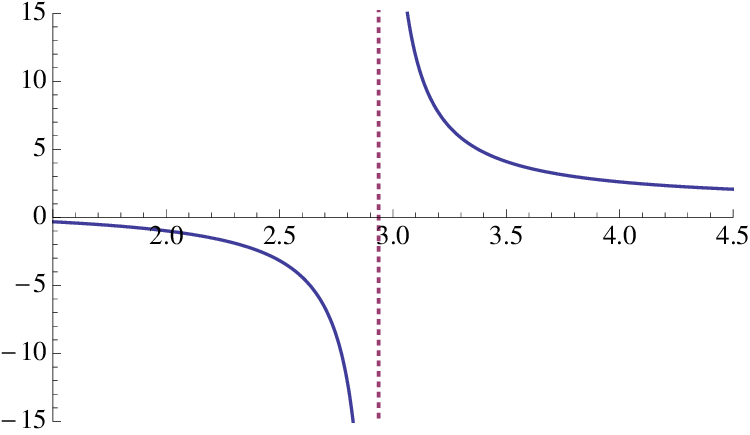}\hfill\hfill
 \includegraphics[width=0.318\textwidth,clip]{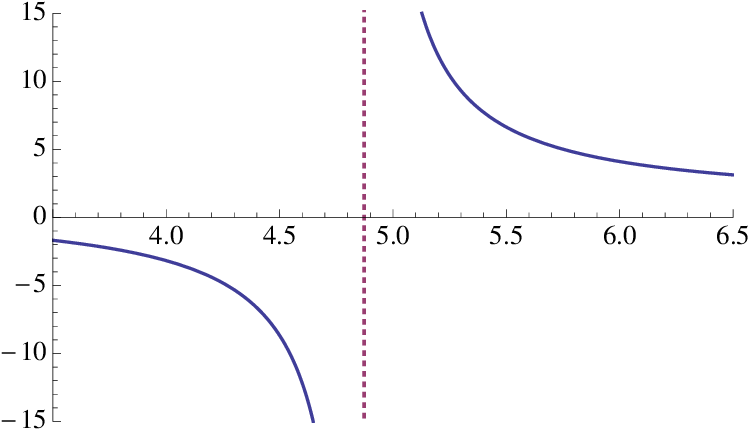}\hfill\hfill
 \includegraphics[width=0.318\textwidth,clip]{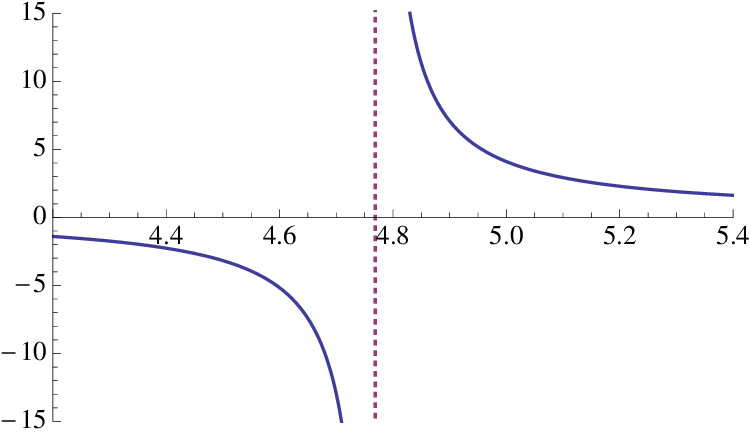}\medskip\\
 \rotatebox{90}{\footnotesize\hspace{8mm}$\Re\,\xi^*(\0)/(zT)$}
 \includegraphics[width=0.32\textwidth,clip]{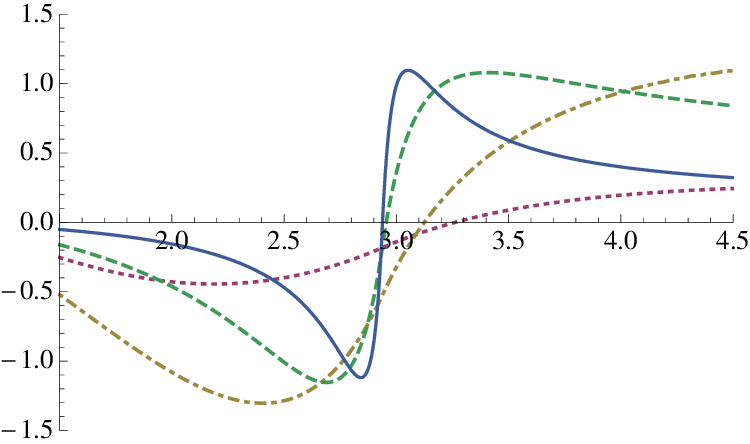}\hfill
 \includegraphics[width=0.32\textwidth,clip]{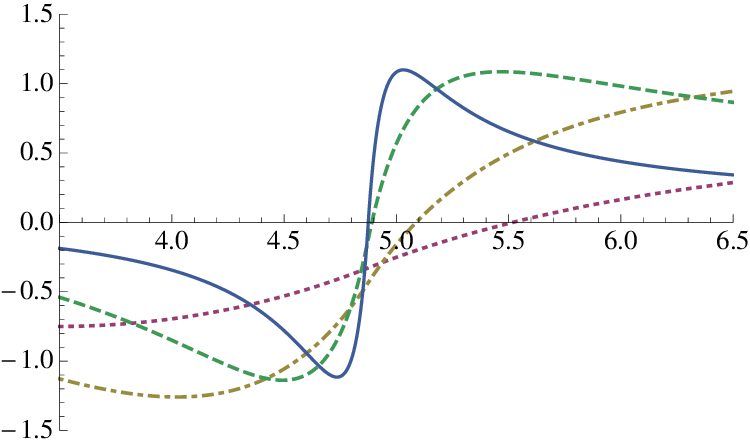}\hfill
 \includegraphics[width=0.32\textwidth,clip]{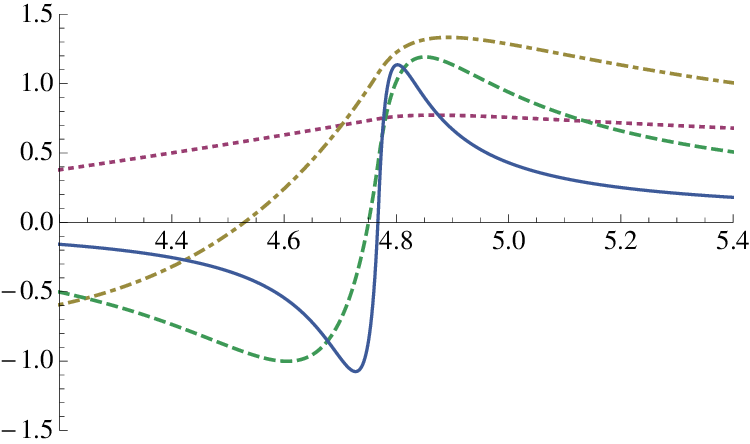}\medskip\\
 \rotatebox{90}{\footnotesize\hspace{9.5mm}$\Im\,\xi^*(\0)/(zT)$}\hfill\hspace{0mm}
 \includegraphics[width=0.312\textwidth,clip]{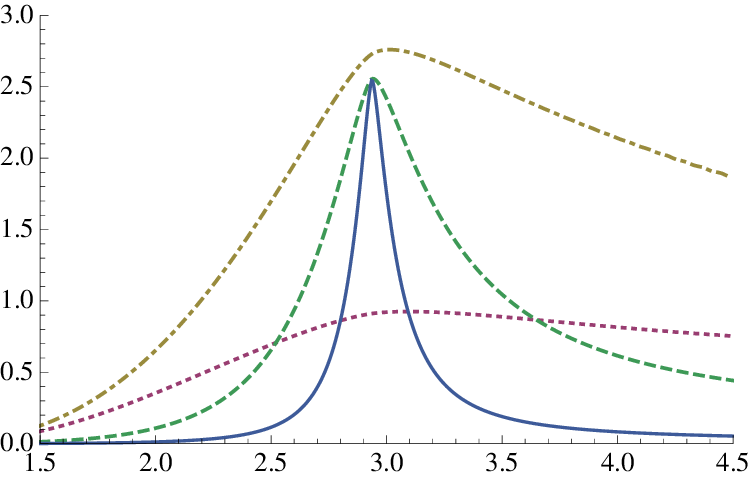}\hfill\hfill
 \includegraphics[width=0.312\textwidth,clip]{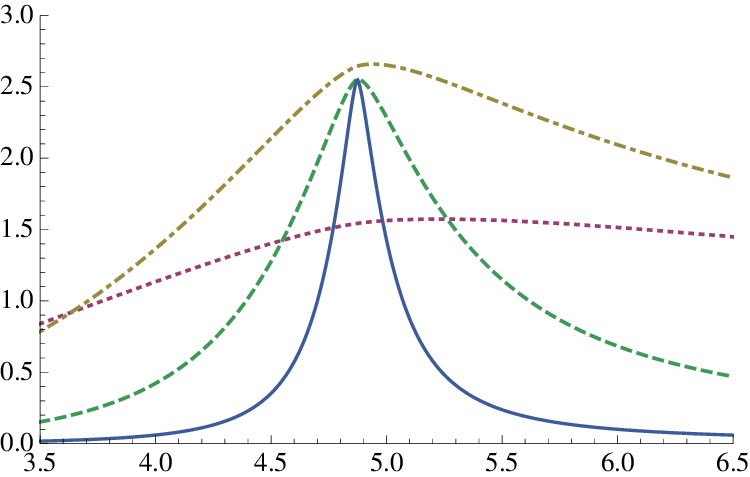}\hfill\hfill
 \includegraphics[width=0.312\textwidth,clip]{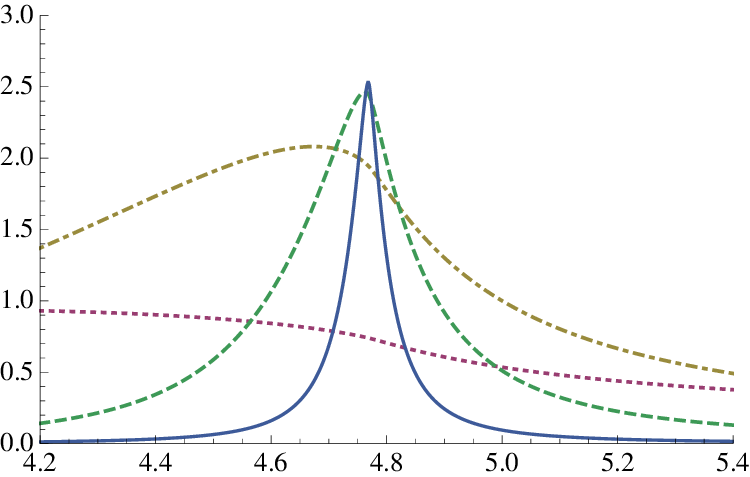}\smallskip\\
 \hspace{7mm}{\footnotesize$J_z/J$}\hspace{51.5mm}
 {\footnotesize$J_z/J$}\hspace{53mm}{\footnotesize$D/J$}
 \caption{Top panels show scattering lengths in
 Eq.~(\ref{eq:scattering-length}) for $S=1/2$ (left), $S=1$ with $D=0$
 (middle) as functions of $J_z/J$, and $S=1$ with $J_z=J$ (right) as a
 function of $D/J$.  The vertical lines indicate the locations of
 two-magnon resonances where $a_s\to\infty$.  Middle and bottom panels
 show real and imaginary parts of $\xi^*(\0)/(zT)$ in
 Eq.~(\ref{eq:magnon_result}) at $T/J=0.01,\,0.1,\,1$, and $10$,
 represented by solid, dashed, dash-dotted, and dotted curves,
 respectively.  \label{fig:magnon}}
\end{figure*}

\subsection{Solution and results}
Our remaining task is to solve the Lippmann-Schwinger equation
(\ref{eq:lippmann-schwinger}) with $i\omega_n=\Delta+i0^+$ and $\p=\0$:
\begin{align}\label{eq:lippmann-schwinger0}
 & \gamma(\r) = 2\cos(\tfrac\q2\cdot\r)
 + \int_{-\pi/a}^{\pi/a}\!\frac{d\k}{(2\pi/a)^3}\cos(\k\cdot\r) \notag\\
 &\times \frac{\sum_{\hat\e}\,[J\cos(\tfrac\q2\cdot\hat\e)
 - J_z\cos(\k\cdot\hat\e)]\,\gamma(\hat\e) - 2D\gamma(\0)}
 {\eq-\varepsilon_{\q/2+\k}-\varepsilon_{\q/2-\k}+i0^+}
\end{align}
to determine $\gamma(\hat\e)$ and $\gamma(\0)$ in the two-magnon
scattering amplitude (\ref{eq:magnon_amplitude}).  In the
low-temperature limit $T\to0$, the integration over $\q$ in the
self-energy (\ref{eq:magnon_self-energy}) is dominated by the region
$\q\simeq\0$ because of the Boltzmann factor.  In this case, by
expanding the right-hand side of Eq.~(\ref{eq:lippmann-schwinger0}) up
to $O(\q)$, we find $\gamma(\hat\x)=\gamma(\hat\y)=\gamma(\hat\z)$ and
the integration over $\k$ can be performed analytically.  Then, by
substituting the obtained analytical solutions for $\gamma(\hat\e)$ and
$\gamma(\0)$ into Eq.~(\ref{eq:magnon_amplitude}), we find that the
two-magnon scattering amplitude takes the same form as that of bosons in
Eq.~(\ref{eq:boson_amplitude}):
\begin{align}
 \lim_{\q\to\0}\Gamma(\Delta+i0^+,\0;\q)
 &= \frac1{a^3}\frac{8\pi/m}{1/a_s+i|\q|/2} \notag\\
 &= \frac1{a^3}F(\Delta+i0^+,\0;\q),
\end{align}
where $1/m=2SJa^2$ is the inverse effective mass of magnons and
\begin{align}\label{eq:scattering-length}
 \frac{a_s}{a} = \frac{\frac3{2\pi}[1-\frac{D}{3J}-\frac{J_z}{J}(1-\frac{D}{6SJ})]}
 {2S-1+\frac{J_z}{J}(1-\frac{D}{6SJ})+3W[1-\frac{D}{3J}-\frac{J_z}{J}(1-\frac{D}{6SJ})]}
\end{align}
is the scattering length between magnons with
\begin{align}
 W &\equiv \int_{-\pi/a}^{\pi/a}\frac{d\k}{(2\pi/a)^3}
 \frac2{\sum_{\hat\e}[1-\cos(\k\cdot\hat\e)]} \notag\\
 &= \frac{\sqrt6}{96\pi^3}\,\Gamma\!\left(\frac1{24}\right)\Gamma\!\left(\frac5{24}\right)
 \Gamma\!\left(\frac7{24}\right)\Gamma\!\left(\frac{11}{24}\right)
\end{align}
being the Watson's triple integral for a simple cubic
lattice.\cite{Watson:1939}  We note that the same scattering length was
obtained in Ref.~\onlinecite{Nishida:2013} with a different approach.

Accordingly, the low-temperature limit of the on-shell self-energy at
$\p=\0$ appearing in Eq.~(\ref{eq:magnon_result}) reduces to that in
Eq.~(\ref{eq:boson_on-shell}),
\begin{subequations}\label{eq:universal}
\begin{align}
 \lim_{T\to0}\xi(\0) = \sigma(\0),
\end{align}
and thus the line shape of the single-magnon peak is described by the
universal formula
\begin{align}
 \lim_{T\to0}I_\mathrm{peak}(\omega) = \frac\omega2SA_\mathrm{peak}(\omega,\0),
\end{align}
\end{subequations}
where $A_\mathrm{peak}(\omega,\p)$ is the single-particle spectral
function of bosons obtained in the previous section as
Eq.~(\ref{eq:boson_result}).  This result is actually expected because
the magnon gas at low temperature is so dilute that the system becomes
independent of microscopic details and thus described by only a few
low-energy parameters such as $m$ and $a_s$ [additionally $\kappa_*$ at
$O(z^2)$].  Therefore, it is possible to extract the scattering length
between magnons by fitting the universal formula (\ref{eq:universal}) to
the experimentally measured temperature dependence of the line shape of
the single-magnon peak.

Away from the low-temperature limit, the line shape
(\ref{eq:magnon_result}) is model dependent and has to be evaluated
numerically by solving Eq.~(\ref{eq:lippmann-schwinger0}) for general
$\q$.  The frequency shift and the linewidth of the single-magnon peak
are given by the real and imaginary parts of $\xi^*(\0)$, respectively,
and the corresponding normalized quantity $\xi^*(\0)/(zT)\sim O(1)$ is
plotted in Fig.~\ref{fig:magnon}.  For demonstration, we choose three
distinct cases where $S=1/2$, $S=1$ with $D=0$ as functions of $J_z/J$,
and $S=1$ with $J_z=J$ as a function of $D/J$ at four different
temperatures, $T/J=0.01,\,0.1,\,1,\,10$.  Figure~\ref{fig:magnon} also
displays the corresponding scattering length
(\ref{eq:scattering-length}) which indicates that the two-magnon
resonances $a_s\to\infty$ are located at $J_z/J=2.94$, $J_z/J=4.87$, and
$D/J=4.77$, respectively, where magnons interact strongly.  We find that
the line shape of the single-magnon peak is well described by the
universal formula (\ref{eq:universal}) at low temperature $T<J$ and thus
the frequency shift changes its sign and the linewidth reaches its
maximum across the two-magnon resonance.  Such characteristic behaviors
become sharper with decreasing temperature and can be seen moderately
even at intermediate temperature $T\simeq J$, while they disappear at
higher temperature $T>J$.

\section{Conclusion and discussion \label{sec:summary}}
In this paper, we studied ESR in a dilute magnon gas that is realized in
a ferromagnetic spin system at low temperature.  We developed the
quantum cluster expansion up to $O(z)$ which is determined by the
two-magnon physics and showed that the frequency shift of the
single-magnon peak changes its sign and the linewidth reaches its
maximum across a scattering resonance between magnons.  Such
characteristic behaviors are universal and can be used to experimentally
locate the two-magnon resonance when an external parameter such as
pressure is varied.  Future achievement of the two-magnon resonance may
have an impact comparable to the Feshbach resonance in ultracold atoms
and will open up a rich variety of strongly correlated physics such as
the recently proposed Efimov effect in quantum
magnets.\cite{Nishida:2013}

It is straightforward in principle to continue this systematic expansion
to include higher-order corrections.  In particular, the $O(z^2)$ term
involves the three-magnon physics and thus it is possible to probe the
Efimov effect with ESR.  When the system comes across the critical
coupling where an Efimov state of three magnons emerges from the
scattering threshold, the linewidth of the single-magnon peak as a
function of the external parameter is expected to show an additional
peak structure caused by the three-magnon resonance on either side of
the two-magnon resonance.  This feature is in analogy with ultracold
atom experiments where an atom loss peak caused by the three-atom or
atom-dimer resonance has been used as a signature of the emergence of an
Efimov trimer.\cite{Kraemer:2006,Knoop:2009}  Similarly, a pair of
universal four-magnon states associated with every Efimov
state\cite{Hammer:2007,Stecher:2009} may be observed with ESR through an
additional peak structure in the linewidth at $O(z^3)$ caused by the
four-magnon resonance.\cite{Ferlaino:2009,Ferlaino:2010}  Besides such
characteristic behaviors in the single-magnon peak, we expect an
additional structure in the ESR spectrum at $\omega<\Delta+E_N$ when $N$
magnons form a bound state with binding energy $E_N<0$.  This structure
shows a threshold singularity at $\omega=\Delta+E_N$ as
$I(\omega)\sim z^{N-1}(\Delta+E_N-\omega)^{(3N-5)/2}$, which may be used
to measure binding energies of magnon Efimov states.  Therefore, ESR is
a powerful experimental technique to investigate the interaction among
magnons and their spectrum.

So far we considered ferromagnetic spin systems where the single-magnon
dispersion relation has a minimum at zero momentum.  On the other hand,
it is also possible to induce the two-magnon resonance and thus the
magnon Efimov effect in spin systems with antiferromagnetic or
frustrated exchange couplings where the single-magnon dispersion
relation has a minimum at nonzero
momentum.\cite{Nishida:2013,Ueda:2009,Ueda:2013}  In these cases,
however, the sharp signature of magnon scattering resonances discussed
in this paper will not appear in ESR because it measures the spectrum
only at zero momentum.  Therefore, different experimental techniques
such as inelastic neutron scattering that can scan momentum space should
be used to probe magnon scattering resonances.  The quantum cluster
expansion developed in this paper will be useful to compute any other
physical observables in a dilute magnon gas.

\acknowledgments
The author thanks C.~D.~Batista and Y.~Kato and acknowledges many
valuable discussions during his visit to RIKEN and YITP in the fall of
2012.  This work was supported by a LANL Oppenheimer Fellowship and JSPS
KAKENHI Grant Number 25887020.


\begin{thebibliography}{99}

\bibitem{Chin:2010}
  C.~Chin, R.~Grimm, P.~Julienne, and E.~Tiesinga,
  Rev.\ Mod.\ Phys.\ {\bf 82}, 1225 (2010).

\bibitem{Regal:2004}
  C.~A.~Regal, M.~Greiner, and D.~S.~Jin,
  Phys.\ Rev.\ Lett.\ {\bf 92}, 040403 (2004).

\bibitem{Zwierlein:2004}
  M.~W.~Zwierlein, C.~A.~Stan, C.~H.~Schunck, S.~M.~F.~Raupach, A.~J.~Kerman, and W.~Ketterle,
  Phys.\ Rev.\ Lett.\ {\bf 92}, 120403 (2004).

\bibitem{Kraemer:2006}
  T.~Kraemer, M.~Mark, P.~Waldburger, J.~G.~Danzl, C.~Chin, B.~Engeser, A.~D.~Lange, K.~Pilch, A.~Jaakkola, H.-C.~N\"agerl, and R.~Grimm,
  Nature (London) {\bf 440}, 315 (2006).

\bibitem{Knoop:2009}
  S.~Knoop, F.~Ferlaino, M.~Mark, M.~Berninger, H.~Sch\"obel, H.-C.~N\"agerl, and R.~Grimm,
  Nat.\ Phys.\ {\bf 5}, 227 (2009).

\bibitem{Nishida:2013}
  Y.~Nishida, Y.~Kato, and C.~D.~Batista,
  Nat.\ Phys.\ {\bf 9}, 93 (2013).

\bibitem{Goodenough:1963}
  J.~B.~Goodenough, {\em Magnetism and the Chemical Bond}
  (Wiley, New York, 1963).

\bibitem{Kawamoto:2001}
  T.~Kawamoto, M.~Tokumoto, H.~Sakamoto, and K.~Mizoguchi,
  J.\ Phys.\ Soc.\ Jpn.\ {\bf 70}, 1892 (2001).

\bibitem{Katsumata:2000}
  K.~Katsumata,
  J.\ Phys.: Condens.\ Matter {\bf 12}, R589 (2000).

\bibitem{Ajiro:2003}
  Y.~Ajiro,
  J.\ Phys.\ Soc.\ Jpn.\ {\bf 72}, 12 (2003).

\bibitem{Kubo:1954}
  R.~Kubo and K.~Tomita,
  J.\ Phys.\ Soc.\ Jpn.\ {\bf 9}, 888 (1954).

\bibitem{Mori:1962}
  H.~Mori and K.~Kawasaki,
  Prog.\ Theor.\ Phys.\ {\bf 27}, 529 (1962);
  Prog.\ Theor.\ Phys.\ {\bf 28}, 971 (1962).

\bibitem{Oshikawa:2002}
  M.~Oshikawa and I.~Affleck,
  Phys.\ Rev.\ B {\bf 65}, 134410 (2002).

\bibitem{Landau-Lifshitz:1980}
  L.~D.~Landau and E.~M.~Lifshitz, {\em Statistical Physics}
  (Butterworth-Heinemann, Oxford, 1980).

\bibitem{Braaten:2006}
  E.~Braaten and H.-W.~Hammer,
  Phys.\ Rep.\ {\bf 428}, 259 (2006).

\bibitem{Brockmann:2012}
  M.~Brockmann, F.~G\"ohmann, M.~Karbach, A.~Kl\"umper, and A.~Wei{\ss}e,
  Phys.\ Rev.\ B {\bf 85}, 134438 (2012).

\bibitem{footnote}
In the absence of the spin anisotropies $J_z=J$ and $D=0$, the
two-magnon scattering amplitude (\ref{eq:magnon_amplitude}) at $\q=\0$
vanishes so that the ESR spectrum shows no frequency shift and
linewidth, which is consistent with the general argument given in
Ref.~\onlinecite{Oshikawa:2002}.  Also, by expanding
Eqs.~(\ref{eq:magnon_amplitude}) and (\ref{eq:lippmann-schwinger}) in
terms of the small spin anisotropies $\epsilon\sim J_z/J-1,D/J\ll1$, it
is easy to find that the frequency shift and the linewidth are
$\Re\,\xi^*(\0)\sim O(\epsilon)$ and $\Im\,\xi^*(\0)\sim O(\epsilon^2)$,
respectively, which are again consistent with the previous
results (Ref.~\onlinecite{Kubo:1954,Mori:1962,Oshikawa:2002}).

\bibitem{Watson:1939}
  G.~N.~Watson,
  Q.\ J.\ Math.\ {\bf 10}, 266 (1939).

\bibitem{Hammer:2007}
  H.-W.~Hammer and L.~Platter,
  Eur.\ Phys.\ J.\ A {\bf 32}, 113 (2007).

\bibitem{Stecher:2009}
  J.~von~Stecher, J.~P.~D'Incao, and C.~H.~Greene,
  Nat.\ Phys.\ {\bf 5}, 417 (2009).

\bibitem{Ferlaino:2009}
  F.~Ferlaino, S.~Knoop, M.~Berninger, W.~Harm, J.~P.~D'Incao, H.-C.~N\"agerl, and R.~Grimm,
  Phys.\ Rev.\ Lett.\ {\bf 102}, 140401 (2009).

\bibitem{Ferlaino:2010}
  F.~Ferlaino and R.~Grimm,
  Physics {\bf 3}, 9 (2010).

\bibitem{Ueda:2009}
  H.~T.~Ueda and K.~Totsuka,
  Phys.\ Rev.\ B {\bf 80}, 014417 (2009).

\bibitem{Ueda:2013}
  H.~T.~Ueda and T.~Momoi,
  Phys.\ Rev.\ B {\bf 87}, 144417 (2013).

\end{thebibliography}
\end{document}